\documentclass[aps,amsmath,amssymb,prl,twocolumn,groupedaddress]{revtex4}
\usepackage{bm}
\usepackage{epsfig}
\usepackage[usenames]{color}

\begin{document}

\title{Influence of Interactions on  Flux and Back-gate Period of Quantum Hall Interferometers}

\author{B.~Rosenow}\thanks{On leave from the Institut f\"ur Theoretische Physik, Universit\"at zu
K\"oln, D-50923, Germany}
\affiliation{Physics Department, Harvard University, Cambridge, MA
02138, USA}
\author{B.~I.~Halperin}
\affiliation{Physics Department, Harvard University, Cambridge, MA
02138, USA} 
\date{November 3, 2006}

\begin{abstract}

In quantum Hall systems with two narrow constrictions, tunneling
between opposite edges can give rise to quantum interference and
Aharonov-Bohm-like oscillations of the conductance. When there is an
integer quantized Hall state within the constrictions, a region
between them, with higher electron density, may form a compressible
island.  Electron-tunneling through this island can lead to residual
transport, modulated by Coulomb-blockade type effects.  We find that
the coupling between the fully occupied lower Landau levels and the
higher-partially occupied level gives rise to flux subperiods smaller
than one flux quantum.  We generalize this scenario to other
geometries and to fractional quantum Hall systems, and compare our
predictions to experiments.

\end{abstract}


\maketitle

Quantum Hall (QH) devices are supposed to be an ideal laboratory for
the study of interference effects, because within a conductance
plateau, the bulk of a sample is insulating and current is confined to
conducting edge states \cite{review}.  Closed interference paths can
be defined with the help of constrictions, which 
mediate tunneling
from one edge to the other. Quantum interference should then manifest
itself in flux- and gate-voltage- dependent conductance oscillations.
For a multiply connected tunneling geometry with flux through a
``hole", the partition function and hence all system properties are
periodic under changing the flux by one quantum $h/e$ \cite{ByYa61},
and in some cases, this is indeed the smallest period.  Subperiods are
allowed, however, and we shall argue below that they will often be
seen in interference experiments, particularly in interacting systems.  Multiple periodicities can
occur in more complicated geometries.

While interference effects in QH systems are well understood for
idealized models, the influence of interactions in more realistic
models has not  been analyzed in detail.  Moreover, the filling
fraction in constriction regions can be different from the bulk
filling fraction \cite{Roddaro+05,Camino+05,CaZhGo05} even giving rise
to fractional QH physics while the bulk is in an integer plateau
\cite{Roddaro+05}.  In this Letter, we study the influence of
interactions on the flux and back-gate \cite{back-gate} periodicity of interference
effects in QH systems with a center island, whose filling fraction is
larger than that of the constriction regions connecting it to the
bulk, see Fig.~1.

Consider a sample with $f_c$ fully occupied Landau levels (LLs) in the
constrictions and an additional partially filled LL in the center
island.  Deviations from the ideal quantized conductance may be caused
by tunneling of electrons through the center island, and when the
tunneling matrix elements are small, this may be strongly modulated by
Coulomb blockade physics of the partially occupied LL.  If one varies
the magnetic field, one also varies the number of electrons contained
in the filled LLs in the center region, 
which couple  electrostatically 
to the partially occupied LL.
  If one
flux quantum is added to the center region, then $f_c$ electrons are
added to the filled LLs, and an equal number must be expelled from the
partially filled level. From this, we shall find a subperiod of
$1/f_c$ flux quanta for the conductance, and a back-gate period
corresponding to one electron charge.  We shall discuss this type of
physics for several geometries, and argue that the theoretically
derived subperiod has already been experimentally observed in integer
QH systems
 \cite{CaZhGo05,ZhCaGo05,vanWees+89,Alphenaar+92}.  
 We also generalize
our findings to the fractional QH regime, and comment on the
interpretation of experiments \cite{Camino+05} in that regime.

\begin{figure}[t]
\includegraphics[width=0.85\linewidth]{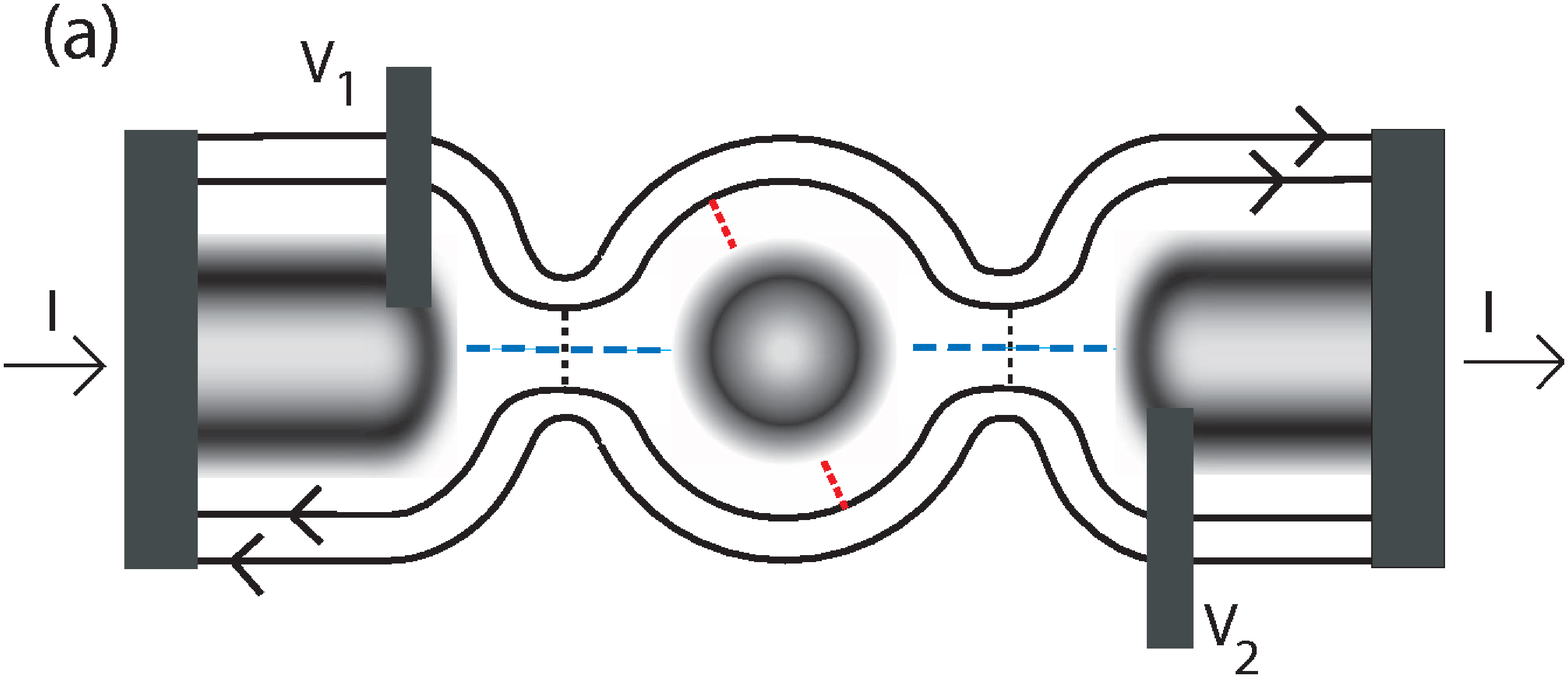}
\includegraphics[width=0.85\linewidth]{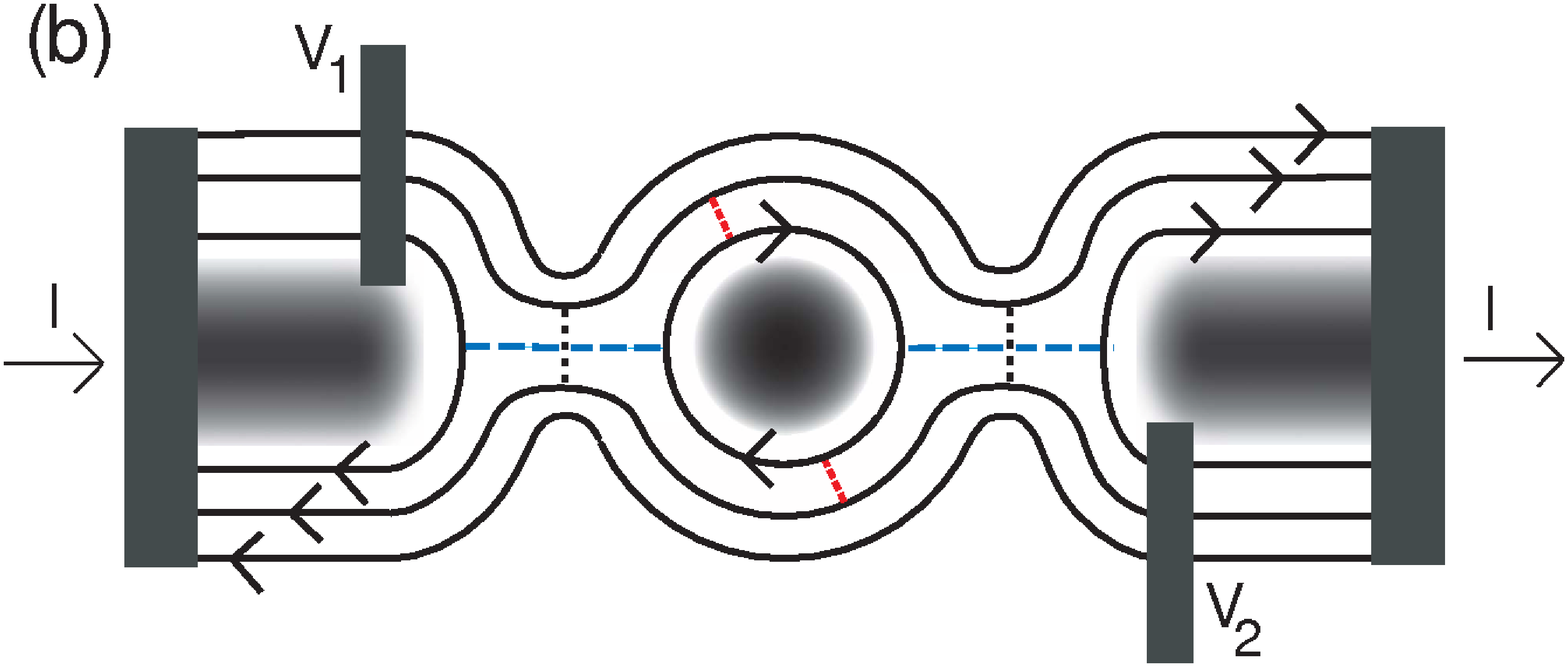}
\caption{Sketch of a QH interferometer with 
(a)  compressible island and
bulk, where light and dark shading indicate regions
of lower and higher conductivity $\sigma_{xx}$. In the lower panel
(b), an additional incompressible region (white) surrounds the
(shaded) compressible regions. Three possible tunneling paths 
indicated by dashed lines: (A) forward
tunneling through the constrictions and center island (blue lines),
(B) backwards tunneling between opposite edge states through the
island (red lines), and (C) backwards tunneling across the
constrictions (black lines).  In the upper panel }
\label{interferometer.fig}
\end{figure}

{\em Description of Geometry:} We consider a Hall bar geometry with
two quantum point contacts (QPCs) and an island between them
\cite{machzender}. 
We assume that the QPCs are sufficiently wide open so that
at zero magnetic field many transverse modes are transmitted.
 Magnetic field and QPC voltages can be tuned such
that the two constrictions are in an integer QH plateau with $f_c$
occupied Landau levels.  In our simplest model
Fig.~\ref{interferometer.fig}a, the actual filling fraction of the
bulk and island regions is assumed to be sufficiently larger than
$f_c$ so that these regions are compressible.  There is both
theoretical \cite{ChShGl92,GeHa94} and experimental
\cite{Ilani+04} support for the idea of spatially extended compressible 
regions. In an alternative picture, 
appropriate for samples with still larger density difference between 
constriction and bulk, we shall assume that
both bulk and island include an incompressible region with $f_b=f_c + 1$  
occupied LLs. (See Fig.~\ref{interferometer.fig}b.)

In the absence of tunneling, both geometries in
Fig.~\ref{interferometer.fig} have a conductance $G\equiv I/ (V_2 -
V_1) = f_c (e^2/h)$.  We consider here several types of tunneling
processes which may modify this conductance: (A.)  Forward tunneling
processes, through the island and the two constrictions, ( dashed blue
lines in Fig.~\ref{interferometer.fig}) can increase the conductance.
Such processes should be particularly important on the low-magnetic
field side of the $f_c$ plateau as the boundaries of the island and of
the compressible regions in the leads become close together.  (B.)
Backscattering processes, which reduce the conductance, can occur
throughout the plateau region, if electrons tunnel from one edge to
the other through the center island (dashed red lines in
Fig.~\ref{interferometer.fig}).  We find that contributions A and B
will be oscillatory functions of magnetic flux or back-gate voltage,
due to Coulomb-blockade-type effects.  (C.) Direct tunneling across
the constrictions (dashed black lines in
Fig.~\ref{interferometer.fig}) can occur, which would again lead to
backscattering and a reduction of the conductance relative to the
plateau value.  This process is most likely to be important on the
high-field side of the plateau.  Process C can lead to oscillatory
conductance if there is quantum interference between particles
tunneling across the two QPCs but the oscillation periods will
generally be different from those of A or B.  In real samples, all
three types of tunneling may occur simultaneously, and it is important
to understand which is the dominant contribution to observed
oscillations.

{\em Tunneling into a compressible island.}  When the tunneling
conductance between island and bulk is much smaller than $e^2/h$,
quantum mechanical delocalization of charge is strongly suppressed,
and the island charge is quantized in units of the electron charge.
If the tunneling amplitudes are sufficiently small, broadening of
levels on the island is due to temperature and not due to the tunnel
coupling.  Conductance across the island is controlled by Coulomb
blockade
 \cite{Lee90,McEuen+92},
  i.e.~tunneling onto the island is only possible
if its electrostatic energy is degenerate, on the scale $k_B T$, with
respect to adding or removing an electron.  The period of conductance
oscillations can be determined by calculating the period of the island
energy with respect to changes in the magnetic field or back-gate
voltage.

Let us choose some value of the magnetic field and back-gate voltage,
and draw a closed curve, of area $A_0$, within the $f_c$
incompressible region, slightly outside the compressible island.  If
we fix the position of this curve, and change the field by a small
amount $\delta B$, the flux through the reference area $A_0$ will
change by $\phi={\delta B A_0/ \Phi_0}$.  Due to the quantized Hall
conductance of the incompressible region, electric fields generated
during the flux change will cause a net charge $f_c \phi e$ to flow
inward through the reference curve, where $e<0$ is the electron
charge.  In addition, an integer number of electrons can hop across
the incompressible region, into or out of the compressible island
changing its charge by $N e$.  Using a back-gate, the positive
background charge in the island area can be increased and $N_{\rm
gate}$ additional electrons can be attracted to the area $A_0$. Hence,
we obtain a total charge imbalance inside the area $A_0$ given by $f_c
e \phi + e N- e N_{\rm gate}$, which leads to a charging energy
%
\begin{eqnarray}
E & = & {e^2 \over 2 C_{i} } ( f_c \phi  \ + \ N - N_{\rm gate})^2 \ \ ,
\label{energy_flux.eq}
\end{eqnarray}
%
where $C_i$ is the capacitance of the island.  Note that the energy
Eq.~(\ref{energy_flux.eq}) returns to its original value when changing
$N$ by -1 and $\phi$ by $1/f_c$, so the magnetic field period is
%
\begin{equation}
\Delta B = {1 \over f_c} {\phi_0 \over A_0} \ .
\label{fieldperiod.eq}
\end{equation}
%

The $f_c$--dependence is caused by the Coulomb repulsion
between electrons, and for $f_c>1$, the magnetic field period
 is strikingly different from the
 period ${\phi_0 / A_0}$ that one would obtain for a
simple Aharonov-Bohm effect in an area $A_0$.  The back-gate \cite{back-gate} period, however,   is one electron charge in the island area 
independent of the filling fraction.
If the field or the back-gate-voltage is changed by an amount which is
too large, we will need to take into account the change in position of
the boundaries between the compressible and incompressible regions,
which will lead in turn to a continuous change in the field period
$\Delta B$ and gate period $\Delta V_G$.  

{\em Calculation of conductance.}  For the calculation of the
fluctuating part of the conductance, one needs to know the addition
energy $\Delta_+^{(N)}$ for adding an electron to the island and the
subtraction energy $\Delta_-^{(N)}$ for removing one.  The conductance
$\delta G$ may then be related, via the fluctuation--dissipation
theorem, to the diffusion rate for motion of electrons into and out of
the island.  One finds $\delta G= M\beta \tilde{D}$, where
%
\begin{eqnarray}
\tilde{D} = {\sum_{N} e^{-\beta E(N)} [f(\Delta_+^{(N)}) + 1 
- f(-\Delta_-^{(N)})] \over \sum_{N} e^{- \beta E(N)}} \  ,
\label{diffusion.eq}
\end{eqnarray}
%
$f(x) = (1+e^{\beta x})^{-1}$ is the Fermi distribution in the
reservoir, with $\beta = 1/(k_B T)$, and $M$ contains the tunneling
matrix elements and other factors, which, for the moment, we treat as
constants. The analysis we have carried out for the forward tunneling
process, A, can also be applied to the backward process B of tunneling
through the island, giving the same flux and gate periods.

\begin{figure}[t]
\includegraphics[width=0.85\linewidth]{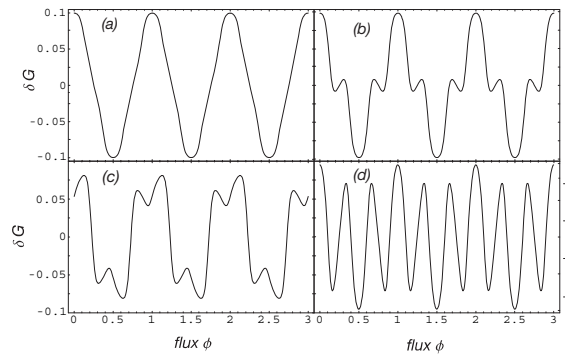}
\caption{Flux dependence of the conductance due to backscattering at
the constrictions (process C) for $\beta e^2/2 C_i = 5$ and $f_c = 4$.
(a) Weak coupling $\Delta_\times / \Delta = 0.1$ between compressible
island and edge.  The period is one flux quantum.  (b) Stronger
coupling $\Delta_\times / \Delta = 0.35$, with zero gate voltage
($N_{\rm gate} = 0)$.  An additional modulation with small amplitude
can be seen. (c) Gate voltage with $N_{\rm gate} = 0.4$, and
$\Delta_\times / \Delta = 0.35$, leads to a splitting of the main
peak.  (d) Strong coupling $\Delta_\times / \Delta = 0.7 $. With one
large and two smaller peaks, one sees an apparent subperiod of $1/(f_c
-1)$ flux quanta.}
\label{fluxperiod.fig}
\end{figure}

{\em Process C: Back--scattering at the constrictions.} If there is
weak backscattering across the constriction regions with $f_c$ fully
occupied LLs (dashed black lines in Fig.~\ref{interferometer.fig}),
there can be interference between paths which scatter at the left and
at the right constriction, respectively.  For the moment, we assume
that the area enclosed by this interference path is the same as the
area of the compressible island, and generalize to different areas
later.  Without the coupling between edge mode and inner island, the
flux-dependence of the interference phase would be $- 2 \pi \phi$.
Due to the Coulomb interaction between the edge and island, a charge
imbalance on the island shifts the edge potential on average by
$\delta V = \Delta_X (f_c \phi + N - N_{\rm gate})$. Here,
$\Delta_X$ is the coupling energy for one extra electron on the
island.  If the total length of the interference path surrounding the
island is $L$, then the effective level spacing along this
interference path is $\Delta = 2 \pi \hbar v/L$, where $v$ is
edge-mode velocity . As the level spacing $\Delta$ corresponds to a
phase shift of $2 \pi$, the potential shift $\delta V$ causes a phase
shift $2 \pi \delta V/ \Delta$. Hence, the total variation in the
conductance is proportional to
%
\begin{equation}
\delta G  \sim  \Big\langle \cos \Big[ - 2 \pi  \phi  +
2 \pi {\Delta_X \over \Delta} (f_c \phi + N - N_{\rm gate})\Big] \Big\rangle_N
\label{oscillation.eq}
\end{equation}
%
The thermal average has to be taken with respect to the number $N$ of
extra electrons on the island and is weighted with a Boltzmann factor
containing the island charging energy Eq.~(\ref{energy_flux.eq}). If
the coupling between island and edge is weak, the resulting flux
period is one flux quantum, while for strong coupling a subperiod of
$1/(f_c -1)$ flux quanta is found.  (See Fig.~\ref{fluxperiod.fig}.)
As the inverse energies $1/\Delta$ and $1/\Delta_\times$ are
proportional to the capacitance per unit length of the edge and the
cross--capacitance between edge and island respectively, we can
estimate the ratio $\Delta_\times/\Delta$ from a purely electrostatic
calculation of the capacitance matrix for a two--dimensional
conducting disk surrounded by (but electrically isolated from) a thin
conducting annulus.  For reasonable input parameters, we find
$\Delta_\times/\Delta \approx 0.35$ (Figs.~\ref{fluxperiod.fig}(b) and
(c)).

{\em Incompressible island.}  The geometry
Fig.~\ref{interferometer.fig}b, with an incompressible region of
filling $f_b=f_c+1$ inside the island is more complicated than that of
a simple compressible island. Here, we envision a relatively narrow
compressible region, or quantum Hall edge state, separating the
incompressible strips at $f_c$ and $f_b$, as well as a compressible
region at the center of the island.  The charging energy will have
contributions from the two compressible regions and the coupling
between them.  For sufficiently strong cross--coupling, the magnetic
field period is again given by Eq.~(\ref{fieldperiod.eq}).  For weak
cross--coupling, the charging of the edge state is approximately
independent from that of the island center, and the field period is
${\phi_0 \over A_0}$. For intermediate coupling strengths, a crossover
between these periods is observed.  If we can ignore the difference
between the overall island area $A_0$ and the area of the inner
compressible region, we find, typically, one large and $f_c -1$
smaller peaks in the period Eq.~(\ref{fieldperiod.eq}), similar to
that shown in Fig.~\ref{fluxperiod.fig}b for a compressible island
with scattering process C .

{\em Several characteristic areas.}  In a more general situation, the
inner compressible region may have an area significantly different
than $A_0$.  For process C, the compressible region may be
significantly smaller than the area between the two QPCs.  Then, the
system is characterized by at least two areas and the resistance will
be, in general, a quasiperiodic function of the magnetic field. If the
areas are commensurate, then the fundamental frequency is given by the
largest common factor of the areas, and a {\em superperiod} could
result.  If the areas are incommensurate and are subject to strong 
electrostatic coupling, 
phase
offsets like $N_{\rm gate}$ in Eq.~(\ref{oscillation.eq}) will vary
with the flux, and both the positions and heights of Coulomb blockade
peaks may appear to vary randomly as a function of magnetic field.
In a Fourier spectrum, several distinct frequencies may be prominent.

{\em What is a compressible region?} Thus far, we have assumed that
the boundaries between compressible and incompressible regions can be
located with some accuracy. The definition of a compressible region
depends, however, on the time-scale of measurement. For Coulomb
blockade energies, we are primarily concerned with equilibrium charge
numbers and charge distributions.  Then, even a very small value of
$\sigma_{xx}$ is sufficient to render a region conducting, or
effectively compressible, and we expect that the incompressible
regions will be very narrow, typically only a few times larger than
the magnetic length.

The concept of Coulomb blockade requires that the number $N$ of excess
electrons inside the area $A_0$ may be treated as an integer. For
this, it is necessary that the total Corbino conductance between the
compressible island and the outside world be small compared to
$e^2/h$.  If $L$ is the perimeter of the compressible island, and $w$
is the width of the incompressible strip, this requires that the
effective value of $\sigma_{xx}$ for the incompressible region must be
small compared to $(w/L)(e^2/h)$.

For an electron to contribute to the conductivity via transport
process A or B, it is necessary that after tunneling into the
compressible island, it can travel half-way around the island edge in
a time comparable to the dwell time on the island. This does not
depend directly on $\sigma_{xx}$; if there is a gradient in the Hall
conductivity $\sigma_{xy}$, an electric charge can move rapidly
perpendicular to the gradient, due to its Coulomb charging energy,
following a contour of constant $\sigma_{xy}$.  However, we may expect
that $\sigma_{xy}$ is very nearly constant in any region where
$\sigma_{xx} \ll e^2/h$, and carriers in the partially filled Landau
level then move only slowly, by hopping processes.  The region useful
for transport, therefore, is only a portion of the compressible strip,
where $\sigma_{xx}$ is large, and the gradient of $\sigma_{xy}$ is
significant, indicated schematically by the dark shaded region in
Fig.~\ref{interferometer.fig}a.  Hence, an electron must not only
tunnel across the incompressible strip, it must also get across the
outer light-shaded region, either by tunneling or by thermally
activated hopping, to reach the dark shaded area. This may cause a
significant decrease in the amplitude of the contribution to the
conductance, and 
affect 
the temperature dependence.  However, the
oscillatory dependence on magnetic field or back-gate voltage should
still be determined by the area $A_0$ of a curve embedded in the
narrow incompressible region.

By contrast, Aharonov-Bohm oscillations due to Process C require fast
transport along the edges, so that there can be quantum interference
between the two constrictions.  For such fast processes, we may
consider that the incompressible regions are broad and the
compressible regions are narrow; i.e., we may treat them as narrow
edge states. Thus, the area which determines the flux $\phi$ for the
interference process (e.g., first term in (\ref{oscillation.eq})) is
the area enclosed by the indicated edge states between the two
constrictions.

{\em Application to fractional QH systems.}  Consider now a device
with an incompressible {\em fractional} QH state in the constrictions,
with filling fraction $f_c = r/s$, and compressible regions in the
island center and the bulk.  Now we expect that charge can tunnel
across the incompressible region, into or out of the island, in units
of the quasiparticle charge $q=e/s$.  For processes analogous to A and
B above, the charging energy corresponding to
Eq.~(\ref{energy_flux.eq}), is then given by 
\begin{equation} E \ = \
{e^2 \over 2 C_i} \ {1 \over s^2} \Big[ r \phi + N - s {N_{\rm gate} }
\Big]^2 \ , \label{fraccharge.eq} \end{equation}
the integer $N$ denotes the number of charge $q$ quasiparticles that
have hopped onto the island. We then find a subperiod $\phi = {1 \over
r}$, similar to the integer case, but a back-gate period $\Delta N =
1/s$.  If other transport mechanisms are important, or if the island
contains an additional fractional QH state, with filling $f_b > f_c$
the situation becomes more complicated, and multiple periods may be
observed as in the analogous integer cases.

{\em Comparison with experiments.} In the integer QH regime, a
Landau-level dependence of the magnetic field period $\Delta B \sim {1
\over f_c}$, as described by Eq.~(\ref{fieldperiod.eq}), has been seen
in experiments with QPCs defined by etch trenches
\cite{CaZhGo05,ZhCaGo05}.  In an earlier experiment \cite{vanWees+89},
a strong dependence of $\Delta B$ on $f_c$ was found as well but
interpreted in terms of a magnetic-field-dependent island radius.  In
a reanalysis 
\cite{CaZhGo05,Dharma-wardana+92}
 of that experiment, however, it was
pointed out that under the assumption of a magnetic field independent
island radius the data agree with $\Delta B \sim {1 \over f_c}$ as
well.

Recently, interference in a fractional QH system was studied
experimentally \cite{Camino+05}. A flux period $\Delta \phi = 5$ and a
back-gate period $\Delta N_{\rm gate} =2$ was observed in a regime
where the bulk was believed to have $f_b=2/5$ and the constrictions
$f_c=1/3$.  The models considered in our paper do not easily explain
these observations.

{\em Conclusions.} A realistic modeling of QH interferometers should
take into account the filling-fraction difference between
constrictions and bulk.  Due to the requirement of charge neutrality,
the interaction between fully occupied lower and the partially
occupied higher Landau levels can give rise to flux subperiods in
quantum Hall interferometers.  Comparison with experiments in the
integer QH regime support our findings.

{\em Acknowledgments.} The authors thank A.~Yacoby, V.~Goldman,
M.~Kastner, C.~Marcus, J.~Miller, A.~Stern, and D.~Zumbuhl for
important discussions.  Work was supported by NSF grant DMR 05-41988,
and by a Heisenberg stipend of DFG for B.~R.


\begin{thebibliography}{10}
\bibitem{back-gate} We assume  spatially uniform capacitive coupling 
between an idealized back gate and charge carriers.
\end{thebibliography}
\end{document}